\documentclass[preprint,12pt]{aastex}

\slugcomment{}

\shorttitle{Risaliti et al.}
\shortauthors{AGNs contribution to FIR background}

\begin{document}

\title{The contribution of Quasars to the Far Infrared Background}

\author{G. Risaliti\altaffilmark{1,2}, M. Elvis\altaffilmark{1}, R. Gilli\altaffilmark{2,3}}



\altaffiltext{1}{Harvard-Smithsonian Center for Astrophysics, 60 Garden Street,
Cambridge, MA 02138}
\altaffiltext{2}{Osservatorio Astrofisico di Arcetri,  Largo E. Fermi 5, I-50125 Firenze, Italy}
\altaffiltext{3}{Dept. of Physics and Astronomy, The Johns Hopkins University, Baltimore, MD 21218}

\begin{abstract}
Recent observational results obtained with SCUBA, COBE and ISO have greatly
improved our knowledge of the infrared and sub-mm  background radiation.
These limits become constraining given the realization that most AGNs are
heavily obscured and must reradiate strongly in the IR/sub-mm.
Here we predict the
contribution of AGNs to the IR/sub-mm background, starting from
measurements of the hard X-ray background. We show that an
application of what we know of AGN Spectral Energy Distributions (SEDs) and the IR background
requires that a significant fraction of the
10-150~$\mu$m background comes from AGNs. This conclusion can only be avoided
if obscured AGNs are intrinsically brighter in the X-rays (with respect to the optical-UV) than
unobscured AGNs, contrary to ``unified schemes'' for AGNs,
or have a dust to gas ratio much lower ($\le 0.1$) than Galactic.
We show that these
results are rather robust and not strongly dependent on the
details of the modeling.
\end{abstract}


\keywords{galaxies: active --- X-rays: diffuse background --- infrared: galaxies}

\newpage
\section{Introduction}

The total energy
released by AGNs, as inferred from the X-ray background intensity at 30 keV, and using
a reasonable bolometric correction, is a significant fraction of the total luminosity of the
Universe
(Fabian \& Iwasawa 1999, hereafter FI99, Elvis,  Risaliti \& Zamorani 2001, hereafter ERZ01).
ERZ01 estimate  that
the total luminosity emitted by AGNs  cannot be lower than 6\% of the total, and could well be
15-20$\%$.
This is a consequence of the large fraction, $\sim$80\%, of AGNs that are heavily obscured.
Heavy obscuration requires re-emission of the absorbed energy and
suggests that the contribution of AGNs to the far infrared background (FIRB) could also be large.
In this work we investigate the contribution of AGNs to the infrared background,
using the latest observational data. Furthermore, the total luminosity emitted by AGNs can be compared with
the mass density of supermassive black holes in order to investigate the
average efficiency of accretion in converting mass into energy. This second important issue is discussed in the
companion  paper ERZ01.

The X-ray spectrum of a type 1 (broad lined) AGN in the 2-30 keV energy range is well
described by a power law with photon index $\sim$ 1.8-2 (Nandra \& Pounds 1994, Nandra et al. 1997).
If the X-ray background is
made by AGNs, then the only way to reproduce
the observed XRB spectrum is by assuming a large number of heavily obscured
AGNs as the main contributors to the XRB, as first suggested by Setti \& Woltjer
(1989). Recent synthesis models that were able to reproduce the XRB spectrum
(Comastri et al. 1995, Gilli, Salvati \& Hasinger 2001, hereafter G01) predict that only 10-20\% of the
XRB at 30 keV is due to type 1 AGNs, while 80\%-90\% of all AGNs are
heavily obscured, type 2 objects.
We can then use these synthesis models to predict the contribution of AGNs
to the FIRB, using an average FIR SED instead of an X-ray SED.

To make an useful estimate would seem to require some knowledge of the FIR SED
of obscured AGN.
Two further uncertainties concern the evolution with redshift of obscured
AGNs, which is still debated, and is probably different from that of unobscured AGNs
(G01), and the column density distribution of these objects at high z. 
Finally, the spectral shape of the FIRB between 10 and 200~$\mu$m
in unknown.
However, we show in this paper that we can draw several important conclusions despite these
uncertainties.


\section{Measurements of the FIRB and Quasar Source Counts}

Here we review
the main observational constraints on the 10~$\mu$m-1 mm background.

{\em 850~$\mu$m}: Deep {\it Chandra} and SCUBA observations of blank fields showed that the
sources making the bulk of the XRB in the 2-10 keV band and those making the 850~$\mu$m background
are unrelated (Severgnini et al. 2000, Fabian et al. 2000).
As a consequence, AGNs are expected to contribute little ($<$ 10\%)
to the sub-mm background. We note however that the SCUBA sources could
also be powered by completely Compton-thick AGNs. This would imply that
the fraction of Compton-thick AGNs at high redshift would be much higher
than in the local Universe. This scenario would strengthen our conclusions,
as we will briefly discuss below.

{\em 140~$\mu$m, 240~$\mu$m}: COBE/DIRBE measurements of the diffuse background
at 140~$\mu$m and 240~$\mu$m (Lagache et al. 1999)  give
$\nu I_{\nu} (140~\mu$m) = 15$\pm$6 nW sr$^{-1}$ m$^{-2}$,
and $\nu I_{\nu} (240~\mu$m) = 11$\pm$2 nW sr$^{-1}$ m$^{-2}$. This result
is somewhat controversial, because of the difficulties in taking into account
the foreground emission. Another group (Fixsen et al. 1998)
finds results higher by 40\%.

{\em 90~$\mu$m}: There is no direct measurement of the background at this wavelength.
The best data at  90~$\mu$m are the recent ISOPHOT source counts (Efstathiou et al. 2000).
However, at the minimum flux detectable at this wavelength (150 mJy) 
the integrated intensity, $\Psi_{90}$, derived from the current logN-logS,
is not constraining, being only a small fraction of the background ($\sim$ 5\%).

{\em 60~$\mu$m}: An analysis of COBE/DIRBE data, performed by Finkbeiner et al. (2000)
gives a 60~$\mu$m background emission of 28$\pm 9$ nW m$^{-2}$ sr$^{-1}$.
However, this measurement is not compatible with the upper limit imposed from
the TeV emission of blazars (Renault et al. 2000). Therefore, there is either another unknown
foreground source to be removed, or the limits estimated from the TeV emission of blazars are
underestimated, due to the poor knowledge of the quasar SED in the TeV range. 
However, we can regard
this value as an upper limit.

{\em 15~$\mu$m}: The ISOCAM survey at 15~$\mu$m provided a logN-logS with S $> 100~\mu$Jy
(Elbaz et al. 1999) and, since in this case the logN-logS curve flattens below the Euclidean value at faint fluxes,
 is likely to have resolved a significant fraction
of the 15~$\mu$m background. If we extrapolate the faint logN-logS to zero flux and calculate the contribution
to the background of the extrapolated part, we find that it is only a few percent
of the contribution from the resolved sources. 

{\em 1-50~$\mu$m}: Observations of blazars in the TeV range give upper limits on the infrared
background from $\sim 1~\mu$m to $\sim 50-60~\mu$m, deduced from the lack
of a cutoff in their TeV spectra which would be produced by Compton scattering of TeV photons
off the IR background (Stanev \& Franceschini 1998).

Putting together these constraints, in Fig. 1 we plot the observational limits
discussed above, and  the arbitrary analytic fit of Fixsen et al. 1998 (thick line)
for the FIRB SED. The plotted quantity, $\Psi$, is defined as $E\times f(E)$, where $E$ is
the energy and f(E) the monochromatic flux density. 
Integrating gives $\Psi_{\rm IRB} = 26$ nW m$^{-2}$ sr$^{-1}$. For comparison,
the $\Psi_{\rm IR,AGN}$ derived for the absorbed AGN radiation by ERZ01 is $\Psi_{\rm IR,AGN}$
= (3.6-6.5) nW m$^{-2}$ sr$^{-1}$.

\section{Constraints on X-ray obscured AGNs from FIR measurements}

Here we compare the observational
constraints on the FIRB summarized above with a quantitative synthesis model

obtained from the code of G01,
that reproduces both the XRB spectrum and the
counts in the hard X-ray surveys of BeppoSAX (Giommi et al. 2000) and ASCA (Akiyama et al.
1999). In this model there are two X-ray SEDs used as input: one is the average SED of unobscured AGNs,
the other is obtained as a weighted average of the SEDs of obscured AGNs with 5 different value of
absorbing column density (for further details, we refer to G01). The XRB spectrum is obtained
integrating these SEDs using the X-ray luminosity functions of Miyaji et al. (2000), leaving only the ratio
between obscured and unobscured AGNs as a free parameter. The best agreement with the
XRB spectrum and source counts is obtained assuming an increase of the number ratio, R, between type 2
and type 1 AGNs from the local R$\simeq$4, to R$\simeq$10 at z$\simeq$1.3.

We used this model, only substituting FIR SEDs in place of the X-ray SEDs.
For type 1 AGNs we assumed the SED
of Elvis et al. (1994) (the normalization was re-scaled from the 2 keV value using $\alpha_{OX}=1.55$ and
$\alpha_{OX}=1.68$\footnote{$\alpha_{OX}$ is defined as the slope of a power law 
connecting the flux density at 2500\AA~ with the flux density at 2 keV, 
as explained below and in ERZ01}.
For type 2 AGNs we assumed a FIR SED with several shapes, in order to be consistent with the observational
data and to investigate how the final results of our analysis depend on the details of
the SED. We discuss these issues below. However  we note that
the integrated intensity obtained with this model
is (obviously) independent of the shape of the SEDs.
We also note that the
details of the redshift distribution of absorbed sources
do not affect much our results, since the ratio between the total absorbed and
transmitted flux is fixed by the slope of the XRB between 2 keV and 30 keV (ERZ01).
The baseline value
we obtained is:
$  \Psi_{base} = 3.8~(C_{\rm BOL}/C_{\rm PG})~{\rm nW~sr}^{-1}~{\rm m}^{-2}$
where C$_{BOL}$ is the  2 keV-to-bolometric correction and C$_{\rm PG}$ is the bolometric correction
for local (z$<$0.4) PG quasars (i.e. assuming  $\alpha_{OX}=1.55$). C$_{\rm PG}$ can be regarded as a lower limit for the real
bolometric correction (ERZ01). Assuming a redshift evolution of  $\alpha_{OX}$ to    $\alpha_{OX}=1.68$ at z=2
as estimated by Yuan et al. (1998), gives C$_{\rm BOL} \sim 2.2  C_{\rm PG}$.
The value we obtain is $\sim$ 2-4 times higher
than the value obtained by FI99 ($\Psi_{\rm FI99}=2$ ${\rm nW~sr}^{-1}~{\rm m}^{-2}$). The fraction of
the total infrared background made by AGN is $\Psi_{base}/\Psi_{IRB}\sim$ 15\%$\times (C_{\rm BOL}/C_{\rm PG}$)

We used several FIR SEDs for absorbed AGNs.
We first assumed a simple, single temperature, blackbody  in the 10-1000~$\mu$m band. The results
are plotted in Fig. 1 for three different temperatures. The normalizations are
determined only by the X-ray data and the bolometric correction discussed above, not by the FIR data.
The allowed temperature range is constrained by the SCUBA data of Severgnini et al. (2000):
average temperatures lower than 120 K overproduce the AGN contribution to the FIRB at 850~$\mu$m.
We repeated the calculation at 120 K allowing for redshift evolution to lower X-ray loudness
(i.e. $\alpha_{OX}$=1.68 at z=2, ERZ01, Yuan et al. 1998). The result
is to shift the curve higher (the upper bound of the shaded region in Fig. 1).
The true bolometric correction then lies
within the shaded region.

A more realistic, multi-temperature, blackbody SED does not change this result significantly. To
investigate the effect of multi-temperature FIR SEDs we held the lowest temperature at 80 K and added
two components with temperatures of 120 K and 160 K. The result of the model is a curve indistinguishable
from that obtained with a single temperature SED of the appropriate temperature ($\sim 110$ K).
This constancy is due to the dominance of redshift dilution, as explained  below.

Strong line emission is a major ISM coolant (Neufeld et al. 1995, Melnick et al. 2000).
Given the poor sampling of the FIRB,
perhaps such sharp line features could be present and so void the limits derived from
broad continuum models.
To test this possibility we assumed a blackbody SED with T=120 K, plus a narrow
emission line at 15~$\mu$m that accounts for $\sim$ half of the total intensity.
This case (which is not intended to refer to a physically realistic model)
is also plotted in Fig. 1 (dashed line) and shows the strong effect of the dilution in redshift:
the strong, narrow feature is broadened over the 15-90~$\mu$m range (the
cut-off at 90~$\mu$m is due to the model cut-off at z=5), and the shape of the final curve
is not markedly different from those obtained with thermal SEDs. Since the feature
we assumed is an extreme case, we conclude that the details of the average
SED of obscured AGNs do not affect our results, only the mean temperature being important.
This ``redshift dilution'' is analogous to the dilution of
the iron K$\alpha$ line in the spectrum of the hard X-ray background (Gilli et al.
1999). Given the independence of the main results from the details of the SED,
in the following we discuss the consequences of our results using only thermal SEDs,
without losing generality. The temperatures we use can be regarded as ``effective''
temperatures, T$_{eff}$, of the actual, unknown SED.

We note that our choice of wavelength range is rather conservative in that it
assumes that the dust is optically thin at all infrared wavelengths.
Indeed, if we assume a standard dust-to-gas ratio, we expect that in
objects with N$_H >$ a few $10^{22}$ cm$^{-2}$ (which constitute more than half of the
local AGN population, Risaliti et al. 1999), the emission
in the 10-20~$\mu$m is absorbed and reradiated at longer wavelengths.
For objects with N$_H >$ a few $10^{23}$ cm$^{-2}$ absorption is still
significant at 25~$\mu$m. If we take this into account, we must
narrow the range of the re-emission, thus increasing the specific intensity
at wavelengths longer than 30~$\mu$m. We will return to this problem
in the next section.

Finally, another possible increase to the AGN contribution to FIRB could
come from a population of Compton-thick AGN, which make no contribution
to the X-ray background, but re-radiate all the intrinsic emission in the far infrared.
In our model we assume that the fraction of such objects is 20\% of
absorbed AGNs, which is a lower limit for nearby low-luminosity Seyfert 2s
(Risaliti et al. 1999). 
Assuming, for example, a fraction of 50\%, the contribution
of AGNs to the 100 $\mu$m background would be probably higher than 50\% (depending
on the SED of these objects).


\section{Discussion}
Our simple argument shows that despite the lack of knowledge of the background
radiation over quite wide wavelength intervals in the far infrared, we can use
the observational
data on the AGN SED  and synthesis models for the X-ray background to
make interesting predictions on the AGNs contribution to the 
FIRB.

Several works have recently computed the AGN contribution to the FIR/sub-mm
background starting from XRB synthesis models (Almaini et al. 1999).
These papers focus mainly on the prediction of the
contribution of AGNs to the SCUBA counts. Trentham \& Blain (2001) already pointed out
that the re-radiation of the primary emission of AGNs cannot be due to
cold ($\sim 40$ K) dust, since this is not consistent with the SCUBA source identifications.
Here we have used a more robust argument, using an improved, higher, AGN bolometric
correction, and deriving the results in a general and only weakly model-dependent way.

From an analysis of Fig. 1 it is clear that (a) the lack of
correlation between the X-ray and 850~$\mu$m background; and (b) the ISOCAM source counts at 15~$\mu$m,
are enough to give new interesting
contraints on the spectrum of the FIRB due to AGNs.
If all our main assumptions are correct, the only way
to obtain agreement with all the observational data is through a
thermal SED with 120 K$<T_{eff}<$ 180 K.

We note that the temperature range can be shifted towards lower values
if we assume a different Spectral Energy Distribution.
In several cases the spectrum of the re-radiated emission is modeled
with a modified thermal function, f(E) $ \propto E^k \times B(E)$, with k$>0$.
We repeated all our calculations assuming k=1. The results are basically
unchanged, with the exception of the temperature interval, which goes from
80 K to 120 K. For comparison, nearby ULIRGs which host an AGN
have an effective temperature between 50 and 80 K, when a SED with k$\sim$1
is assumed. We note however that it is at present extremely difficult
to disentangle the starburst and AGN components even in the most studied
ULIRGs. It is probable that in many sources hosting an AGN, a starburst
contribution  is also important. This component can lower the measured
effective temperature.

The contribution of AGNs to the FIRB strongly affects constraints
on galaxy evolution. In background synthesis models in which the stellar component is dominant,
for example, a ``passive galaxy evolution'' model
cannot reproduce the whole FIRB, unless the stellar initial mass function
is strongly different from the Salpeter IMF and is cut at low masses (Franceschini 2000). The significant
contribution of at least $\sim 15$\% to the FIRB from AGNs indicated by our analysis
alleviates this problem, since the
total energy emitted by stars and reprocessed in the FIR will be significantly lower by a factor at least 1.7.

Alternatively fast galaxy evolution with redshift reproduces the FIR counts in deep surveys
with no contribution of AGNs (Chary \& Elbaz 2001).
In this case, as emphasized by these authors, a contribution of AGNs
to the FIRB higher than $\sim$ 15\%, as we derive, would imply a slower evolution
of the galaxy luminosity functions with redshift.
Assuming T$_{eff} \sim 120 K$ we predict a maximum of the AGN contribution from 20\% to 40\%
at 60~$\mu$m. If the effective temperature, T$_{eff}$, is higher, the wavelength at which
we expect the maximum contribution is lower. In these cases the AGN contribution can
easily be higher than 50\%, and could even reach 100\%.
Again, we remind that  if we
take into account the absorption at 10-30$~\mu$m in the most heavily
absorbed objects, we predict higher emission at 60-100~$\mu$m,
and a lower one at $\lambda < 30~\mu$m.

Our argument is not 100\% watertight: (a) the dust content of the AGNs making the X-ray background
could be significantly lower than currently assumed, or (b) the dust composition
could be different. \\
(a) Several recent studies suggest low dust-to-gas ratio:
Granato et al. (1997) propose a low dust-to-gas ratio in X-ray obscured
AGNs, from the comparison of the infrared data with their models. Spectroscopically selected quasars
suggest the existence of a significant
population of AGNs with a low dust-to-gas ratio
(Risaliti et al. 2001). As a
consequence, the contribution of AGNs to the FIRB estimated in these cases
could be negligible: if we assume that the re-radiation band extends down to wavelengths of
$\sim$ 1~$\mu$m, the FIR emission due to AGNs drop by $\sim 30$\%. The drop could be
much lower if a significant fraction of the optical-UV primary emission is not reprocessed in the
IR. This is for example the case of Broad Absorption Line Quasars, which are heavily absorbed
by gas in the X-rays (Mathur, Elvis \& Singh 1995, Brandt et al. 2000) but little reddened by dust. \\
(b) If dust grains in AGNs are
much larger, on average,  than the standard Galactic composition then
the absorption curve of dust would be flatter, and the absorption efficiency significantly
lower, than for dust with Galactic size distribution (Laor \& Draine 1993). A recent study
of the continuum and line absorption in nearby intermediate AGNs suggest that this
could be the correct scenario for a significant fraction of AGNs (Maiolino et al. 2001a, 2001b).

Finally, we note that the obscured AGNs that make the X-ray
background may have intrinsically higher X-ray to optical ratio.
As a consequence, the bolometric
correction we adopted would be overestimated and the total intensity
reradiated in the far infrared would be correspondingly lower.
This scenario is an alternative to unification models, and  cannot be ruled out,
given our lack of knowledge on the
so-called ``quasar 2s''.

In the near future several new observational opportunities are expected, which will
tightly constrain our model, revealing the properties and the amount of dust of the
AGN contributing to the FIRB: (a) the new X-ray satellites Chandra and XMM-Newton are already providing
X-ray spectra of high redshift AGNs. From the comparison between the X-ray
and infrared properties of these objects we will be able to have a direct measurement
of the contribution of high redshift AGNs to the FIRB.
Complementary information, at lower redshift will come from the infrared
characterization  of the sources found in the deep ASCA and BeppoSAX surveys,
like HELLAS (Fiore et al. 1999). (b)
Sensitive infrared surveys and identifications, most notably with SIRTF, will test
the predictions in Figure 2. More precisely, SIRTF will allow us to measure the
SEDs of the hard X-ray background sources, thus giving a direct test for
the model proposed here. It will also allow us to study via follow-up of
70~$\mu$m sources, other contributors to the FIRB (like, for example,
Compton thick dusty AGNs).

Our model has two main degrees of freedom, as discussed above:
the effective temperature of the assumed SED, and the total amount of
radiation reprocessed by dust.
The optical identification of the sources found in the ISOCAM survey at 15~$\mu$m
will give a first useful constraint.
The AGN contribution  to the FIRB around $\sim 60~\mu$m is quite stable to changes in
T$_{eff}$. SIRTF will resolve the bulk of the background at 70~$\mu$m, reaching a flux
of $\sim 100~\mu$Jy for point sources, with accurate positions that will enable ready
identifications. SIRTF will also resolve a significant fraction
of the 160~$\mu$m (flux limit $\sim 1$ mJy).
These data will tightly constrain our model:
a high AGN contribution at 15~$\mu$m ($>$50\%) would constrain
the SED effective temperature to around 160-180 K. This would imply that the dust
responsible to the reprocessing of the optical and UV radiation is much warmer than
usually assumed, and that the total amount of dust in AGNs has to be low, with respect
to the gas column densities measured in the X-rays.
Conversely, a low AGN contribution at 15~$\mu$m would imply that the dust is on
average colder or that the UV and optical primary emission is not almost entirely reprocessed
by dust (i.e. there are more dust-free lines of sight than assumed from the X-rays).
The 70~$\mu$m measurement would discriminate between these two scenarios.
Finally, if the AGN contribution to the 70~$\mu$m background is dominant, contrary to our
predictions, this would imply that the fraction of completely Compton-thick AGN is much
higher than we assumed.

Summarizing, in this work we have  combined all the available data
on AGN SEDs and X-ray background models on one side, and on the FIRB on the
other side. We have found, with an approach that is only weakly model-dependent, that
the contribution of AGNs to the infrared background is not negligible,
and could be dominant in the 15-50~$\mu$m band (if, for example, the fraction
of Compton thick AGNs is higher than our conservative estimate of 20\%). 
This has large consequences for
models of galaxy evolution. The only possible
alternatives are that AGNs have quite different SEDs than usually assumed,
or that the dust content or composition in absorbed AGNs is quite different
from Galactic. These quantities will be tested soon with ISO and SIRTF.

\acknowledgments
We are grateful to G. Zamorani for fruitful discussions.
This work was supported in part by NASA grant NAG5-4808.


\clearpage

\figcaption[corr.ps]{Thick continous line: FIRB, obtained using the analytic fit
of Fixsen et al. (1998) in the 1000-120~$\mu$m range, and connecting the 120~$\mu$m
and the 15~$\mu$m points with a powerlaw.
The points at 140 and 240~$\mu$m are from COBE (Lagache et al. 1999),
the point at 15~$\mu$m is our estimation based on ISOCAM logN-logS (Elbaz et al. 1999); the upper
limit at 850~$\mu$m refers to the AGNs contribution to the background, and is estimated
on the basis of the SCUBA results in Severgnini et al. (2000).
The continous line on the top right of the diagram is an upper limit obtained from
TeV spectra of Blazars (Franceschini 2000).
The three dotted lines are the results of our models,
for 3 different dust temperatures, assuming the X-ray to bolometric correction of
local quasars. The shaded region has its lower bound the 120 K SED and uses the bolometric
correction of local quasars, while the upper bound is given by the same SED, but assuming the
bolometric correction corresponding to an observed $\alpha_{OX}$=1.68.
The dashed line is obtained assuming a dust
temperature of T=120 K and a narrow
emission line at 15~$\mu$m that accounts for half of the total intensity.
The thin continous line is obtained assuming a more complex SED, given by the sum of three
thermal SEDs with three different temperatures (160 K, 120 K, 80 K).
\label{fig1}}

\begin{figure}
\plotone{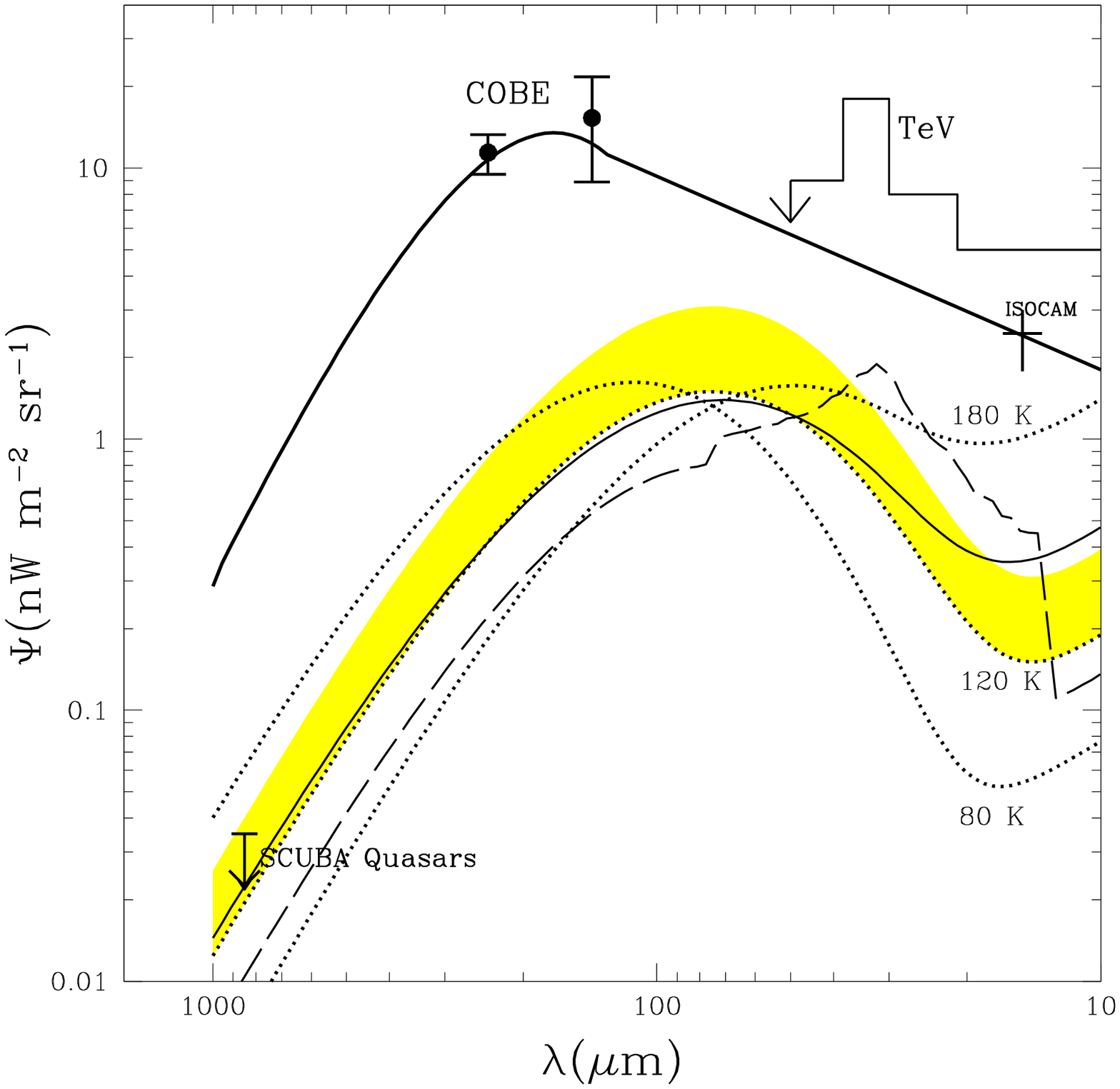}
\end{figure}

\end{document}